\documentclass[conference]{IEEEtran}
\IEEEoverridecommandlockouts
\usepackage{cite}
\usepackage{amsmath,amssymb,amsfonts}
\usepackage{algorithmic}
\usepackage{graphicx}
\usepackage{textcomp}
\usepackage{xcolor}
\usepackage{multirow}
\usepackage[super]{nth}
\def\BibTeX{{\rm B\kern-.05em{\sc i\kern-.025em b}\kern-.08em
    T\kern-.1667em\lower.7ex\hbox{E}\kern-.125emX}}
\begin{document}

\title{Decoding High-level Imagined Speech using Attention-based Deep Neural Networks
\footnote{{\thanks{This research was supported by the Defense Challengeable Future Technology Program of Agency for Defense Development, Republic of Korea. }
}}
}

\author{\IEEEauthorblockN{Dae-Hyeok Lee}
\IEEEauthorblockA{\textit{Dept. Brain and Cognitive Engineering} \\
\textit{Korea University} \\
Seoul, Republic of Korea \\
lee\_dh@korea.ac.kr} \\

\and

\IEEEauthorblockN{Sung-Jin Kim}
\IEEEauthorblockA{\textit{Dept. Artificial Intelligence} \\
\textit{Korea University} \\
Seoul, Republic of Korea \\
s\_j\_kim@korea.ac.kr} \\

\and

\IEEEauthorblockN{Keon-Woo Lee}
\IEEEauthorblockA{\textit{Dept. Brain and Cognitive Engineering} \\
\textit{Korea University} \\
Seoul, Republic of Korea \\
kw-lee@korea.ac.kr}
}

\maketitle

\begin{abstract}
Brain-computer interface (BCI) is the technology that enables the communication between humans and devices by reflecting humans’ status and intentions. When conducting imagined speech, the users imagine the pronunciation as if actually speaking. In the case of decoding imagined speech-based EEG signals, complex task can be conducted more intuitively, but decoding performance is lower than that of other BCI paradigms. We modified our previous model for decoding imagined speech-based EEG signals. Ten subjects participated in the experiment. The average accuracy of our proposed method was 0.5648 ($\pm$0.0197) for classifying four words. In other words, our proposed method has significant strength in learning local features. Hence, we demonstrated the feasibility of decoding imagined speech-based EEG signals with robust performance.
\end{abstract}

\begin{small}
\end{small}

\section{INTRODUCTION}
Brain-computer interface (BCI) is the technology that enables the communication between humans and devices by reflecting humans’ status and intentions \cite{kwak2019error, kwon2019subject}. Non-invasive BCI technology has been focused on owing to low cost and it is no need to undergo surgical implantation \cite{lee2020continuous, cho2021neurograsp, lee2021decoding}. For the past decade, therefore, non-invasive BCI technology has been investigated for controlling external devices \cite{kim2016commanding, lee2018high, jeong2020brain} or early detecting some diseases \cite{zhang2017hybrid, zhang2019strength}.

One of the most attractive issues in the BCI domain is communicating with a drone using the electroencephalogram (EEG) with robust performances \cite{lee2021design, karavas2015effect, lafleur2013quadcopter}. Since EEG signals can reflect humans’ current status and intentions, control of the drone using EEG signals has the advantage of being able to flexibly cope with unexpected situations. Lee \textit{et al}. \cite{lee2021design} designed endogenous BCI paradigms (motor imagery, visual imagery, and imagined speech) for acquiring EEG signals. Various tasks related to controlling the drone swarm were instructed to subjects. Karavas \textit{et al}. \cite{karavas2015effect} investigated the effect of swarm cohesion on the EEG activity of the human supervisor. They showed that brain activity is correlated to swarm cohesion levels, which refers to spreading in the motion of the swarm agents. Lafleur \textit{et al}. \cite{lafleur2013quadcopter} examined the impact that the operation of a real-world device has on subjects' control in comparison to 2-D virtual cursor task. Individual subjects were able to accurately acquire up to 90.5 \% of all valid targets presented.

In the BCI domain, stimulus-based exogenous BCI and imagination-based endogenous BCI exist. Stimulus-based exogenous BCI has high accuracy but has high fatigue of the eye. Also, additional equipment for providing stimulation is required. The disadvantages of exogenous BCI can be solved by using imagination-based endogenous BCI. Imagined speech, one of the endogenous BCI paradigms, is that the users imagine the pronunciation as if actually speaking. Bakhshali \textit{et al}. \cite{bakhshali2020eeg} proposed the method for extracting EEG features of imagined speech effectively. They showed that Riemannian distance extracts features in EEG signals effectively. Nguyen \textit{et al}. \cite{nguyen2017inferring} proposed a novel method based on covariance matrix descriptors, which lie in Riemannian manifold and the relevance vector machines classifier. Their method was shown to outperform other approaches in the field with respect to accuracy and robustness. However, most studies of decoding imagined speech-based EEG signals show low performances and use machine learning algorithms to analyze data. Lee \textit{et al}. \cite{lee2019towards} analyzed imagined speech-based EEG signals in the perspective of its presence, spatial features using the random forest (RF), and the shrinkage regularized linear discriminant analysis. Hern{\'a}ndez-Del-Toro \textit{et al}. \cite{hernandez2021toward} used five feature extraction methods for decoding imagined speech-based EEG signals. These methods were tested in three datasets using RF, support vector machine, k-nearest neighbors, and logistic regression.

In this paper, we modified our previous method \cite{lee2021subject} for increasing the performance of decoding imagined speech-based EEG signals. Our proposed attention-based deep neural networks, called ADNN, could classify various words robustly by adding the attention module. The attention module helps the model learn to focus more on the important part to focus on. In other words, the attention module enables increasing the ability of expression by focusing on important features and suppressing unimportant features. To the best of our knowledge, this study is the first attempt to use the attention module for decoding imagined speech-based EEG signals. The proposed framework has achieved the best performance (0.5648 ($\pm$0.0197)). Hence, we demonstrated the feasibility of classifying the imagined speech-based EEG signals with robust performance.

The rest of this paper is organized as follows. In Section 2, we introduce our experimental design and the structure of the proposed model more in detail. In Section 3, we present the experimental results. In Section 4, we present the directions for decoding imagined speech-based EEG signals and the limitations of our study. Finally, we conclude this paper in Section 5.\\

\section{MATERIALS AND METHODS}
\subsection{Subjects}
A total of ten healthy subjects (S1--S10, ten males, aged 25.8 ($\pm 2.5$)) participated in our experiment. Our experimental environment and protocols were approved by the Institutional Review Board at Korea University (KUIRB-2020-0318-01).

After the subjects were informed about the experimental protocols, they consented according to the Declaration of Helsinki. In addition, we instructed the subjects to get adequate sleep (over seven hr.) and to avoid any alcohol the day before the experiment.\\

\subsection{Experimental Environment}
The signal amplifier (BrainAmp, Brain Products GmbH, Germany) was used for the measurement of subjects' EEG signals. The sampling frequency was set up to 1,000 Hz and a 60 Hz notch filter was applied for removing DC noise. 58 EEG channels were used for acquiring EEG signals, and we placed them on the subjects' scalp according to the international 10/20 system. Also, we measured EOG signals by attaching six electrodes around the subjects' eyes. We used the FCz and FPz channels to reference and ground electrodes, respectively. We set the impedance of all EEG electrodes to lower 10 k$\Omega$ or less by injecting the conductive gel into the subjects' scalp before the acquisition of EEG signals.\\

\subsection{Experimental Paradigm}
The experimental paradigm was designed for acquiring imagined speech-based EEG signals with high quality as shown in Fig. 1. Four words (`/Ba/', `/Ku/', `/He/', and `/Li/') were selected according to Levenshtein distance and the Soundex algorithm. We selected four words based on these criteria since the selection of words with a large difference in each index value is important. Our experimental procedure consisted of three phases. One of four words was displayed for 2 sec., randomly. A fixation cross was provided for 1 sec., and a blank image followed for 2 sec.. We instructed the subjects to perform an imagination when a blank image was displayed. A fixation cross and a blank image were repeated four times per word. After that, a bold fixation cross followed for 3 sec. to eliminate the afterimage of the existing word. We acquired 50 trials per word (a total of 200 trials).\\

%%%%%%%%%%%%%%%%%%%%%%%%%%%%%%%%%%%%%%%%%%%%%%%%%%%%%%%%%%%%%%%%%%%%%%%%%%%%%%%%
\begin{figure}[t!]
\centering
\scriptsize
\centerline{\includegraphics[width=\columnwidth]{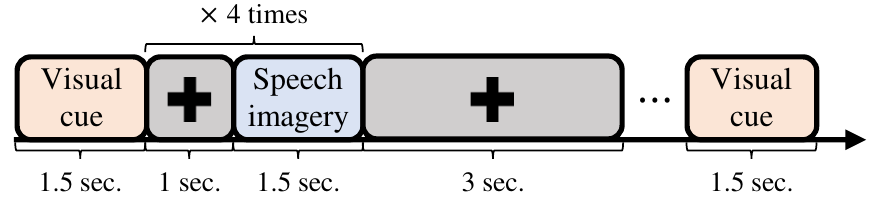}}
\caption{Experimental paradigms for acquiring imagined speech-based EEG signals with high quality.}
\end{figure}
%%%%%%%%%%%%%%%%%%%%%%%%%%%%%%%%%%%%%%%%%%%%%%%%%%%%%%%%%%%%%%%%%%%%%%%%%%%%%%%%

\subsection{Attention-based Deep Neural Networks (ADNN)}
We used EEGNet \cite {lawhern2018eegnet} as a backbone network. The EEGNet is characterized by a small number of parameters, using the depthwise separable convolutional blocks. The network consists of temporal convolution and spatial depth-wise convolution blocks, followed by the depth-wise separable convolutional blocks. Each convolutional block uses a batch norm and dropout, and an ELU function \cite{clevert2015fast} is used as an activation function.

In this study, we modified our previous method \cite{lee2021subject} that can increase the performance of decoding imagined speech-based EEG signals by adding the attention module after the end part of EEGNet \cite{vaswani2017attention, dosovitskiy2020image}. Our proposed method enables increasing the ability of expression by focusing on important features for decoding imagined speech-based EEG signals and suppressing unimportant features as shown in Fig. 2. \\

%%%%%%%%%%%%%%%%%%%%%%%%%%%%%%%%%%%%%%%%%%%%%%%%%%%%%%%%%%%%%%%%%%%%%%%%%%%%%%%%
\begin{figure*}[t!]
\centering
\scriptsize
\centerline{\includegraphics[width=0.90\textwidth, height=0.55\textwidth]{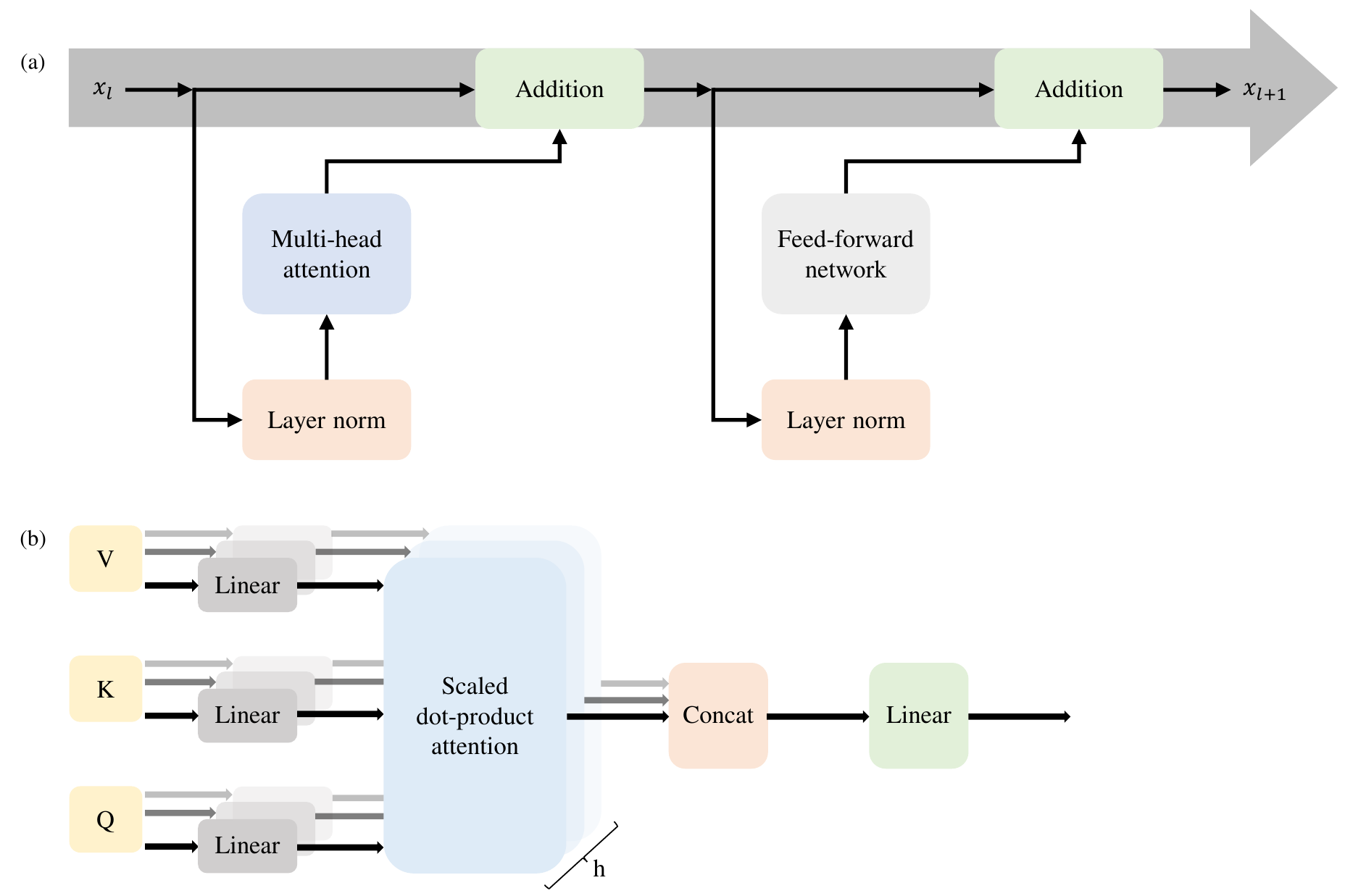}}
\caption{The overall architecture of ADNN including pre-layer normalization and multi-head attention. (a) Pre-layer normalization architecture to train the model with stability, (b) Multi-head attention to train the multiple aspects of data.}
\end{figure*}   
%%%%%%%%%%%%%%%%%%%%%%%%%%%%%%%%%%%%%%%%%%%%%%%%%%%%%%%%%%%%%%%%%%%%%%%%%%%%%%%%

\section{RESULTS AND DISCUSSION}
\subsection{Comparison of Performance with the Conventional Methods}
Table I showed the results of comparing the performances for decoding imagined speech-based EEG signals. We used the power spectral density-support vector machine (PSD-SVM) \cite{zhang2017design} and common spatial pattern-linear discriminant analysis (CSP-LDA) \cite{alimardani2017weighted} as the conventional methods for performance comparison. The PSD-SVM is that used the PSD of the \textit{$\delta$}- (1-4 Hz), \textit{$\theta$}- (4-8 Hz), \textit{$\alpha$}- (8-13 Hz), and \textit{$\beta$}-bands (13-30 Hz) as a feature and used SVM for the classification. The CSP-LDA is that used the CSP algorithm for extracting the features in EEG signals and used LDA for the classification.

Our proposed method represented the highest average decoding accuracy of 0.5648, compared to the performance of the conventional methods. In the performance of our proposed method, S1 showed the highest accuracy of 0.6084, but S6 and S9 showed the lowest accuracy of 0.5472. The method showing the second highest performance is CSP-LDA. In the performance of CSP-LDA, S9 showed the highest performance of 0.5072, while S3 showed the lowest performance of 0.3621. The method showing the lowest performance was PSD-SVM. In terms of PSD-SVM performance, S1 showed the highest performance of 0.4723, while S2 showed the lowest performance of 0.3471. Our proposed method showed the highest average accuracy and the lowest variation. The lowest variation indicated that decoding of imagined speech-based EEG signals was stable across all subjects.

%%%%%%%%%%%%%%%%%%%%%%%%%%%%%%%%%%%%%%%%%%%%%%%%%%%%%%%%%%%%%%%%%%%%%%%%%%%%%%%%
\begin{table}[t!]
\caption{Comparison of performances for the VI classification in the subject-independent task}
\renewcommand{\arraystretch}{1.3}
\tiny
\resizebox{\columnwidth}{!}{
\begin{tabular}{cccc}
\hline
\# of subjects & PSD-SVM \cite{zhang2017design} & CSP-LDA \cite{alimardani2017weighted} & Proposed \\ \hline
S1             & 0.4723  & 0.4944  & 0.6084   \\
S2             & 0.3471  & 0.3840  & 0.5834   \\
S3             & 0.4386  & 0.3621  & 0.5497   \\
S4             & 0.3972  & 0.3967  & 0.5597   \\
S5             & 0.3550  & 0.3735  & 0.5697   \\
S6             & 0.3477  & 0.4070  & 0.5472   \\
S7             & 0.4631  & 0.4736  & 0.5522   \\
S8             & 0.4055  & 0.3742  & 0.5747   \\
S9             & 0.4382  & 0.5072  & 0.5472   \\
S10            & 0.3550  & 0.4291  & 0.5559   \\ \hline
Avg.           & 0.4020  & 0.4202  & 0.5648   \\
Std.           & 0.0491  & 0.0535  & 0.0197   \\ \hline                     
\end{tabular}}
\end{table}
%%%%%%%%%%%%%%%%%%%%%%%%%%%%%%%%%%%%%%%%%%%%%%%%%%%%%%%%%%%%%%%%%%%%%%%%%%%%%%%%

In addition, we conducted the paired \textit{t}‐test with Bonferroni's correction to verify the difference of classification between the conventional methods and ADNN. Initially, we validated the normality and homoscedasticity due to a small number of samples. The normality applying the Shapiro–Wilk test for all methods was satisfied with a null hypothesis, and the assumption of homoscedasticity based on Levene’s test was also satisfied for all models. Our proposed ADNN had the statistically significant difference in performance among all conventional methods ($\textit{p}<0.01$).\\

\subsection{Ablation Study about the Effect of the Attention Module}
Table II indicated the results of the ablation study about the effect of the attention module. To verify the effect of the attention module, we calculated the performance in three kinds of models using `EEGNet’, `EEGNet with SEFE’, and `EEGNet + attention module’. Table II showed that ‘EEGNet + attention module’ improved the performance of EEGNet compared to `EEGNet with SEFE’. Also, after we conducted a paired \textit{t}-test to verify the difference of performance, we showed that `EEGNet + attention module’ showed the statistically significant differences compared to other methods ($\textit{p}<0.01$). From these results, we found that the attention module increases the ability of expression by focusing on important features and suppresses unimportant features. Compared to `EEGNet' and `EEGNet with SEFE', which have significant decoding performance, our proposed method showed the highest performance, proving that our method has significant strength in learning local features.\\

%%%%%%%%%%%%%%%%%%%%%%%%%%%%%%%%%%%%%%%%%%%%%%%%%%%%%%%%%%%%%%%%%%%%%%%%%%%%%%%%
\begin{table}[t!]
\caption{Comparison of performances for the VI classification in the subject-independent task}
\renewcommand{\arraystretch}{1.4}
\scriptsize
\resizebox{\columnwidth}{!}{
\begin{tabular}{cccc}
\hline
\# of subjects & EEGNet \cite{lawhern2018eegnet} & EEGNet with SEFE \cite{lee2021subject} & Proposed \\ \hline
S1             & 0.5642 & 0.5361        & 0.6084   \\
S2             & 0.5254 & 0.4911        & 0.5834   \\
S3             & 0.5142 & 0.5536        & 0.5497   \\
S4             & 0.5254 & 0.5374        & 0.5597   \\
S5             & 0.5067 & 0.5111        & 0.5697   \\
S6             & 0.4854 & 0.5124        & 0.5472   \\
S7             & 0.5004 & 0.4899        & 0.5522   \\
S8             & 0.4929 & 0.5111        & 0.5747   \\
S9             & 0.4929 & 0.5186        & 0.5472   \\
S10            & 0.4979 & 0.5111        & 0.5559   \\ \hline
Avg.           & 0.5106 & 0.5172        & 0.5648   \\
Std.           & 0.0233 & 0.0201        & 0.0197   \\ \hline                    
\end{tabular}}
\end{table}
%%%%%%%%%%%%%%%%%%%%%%%%%%%%%%%%%%%%%%%%%%%%%%%%%%%%%%%%%%%%%%%%%%%%%%%%%%%%%%%%

\section{CONCLUSION AND FUTURE WORKS}
In this study, we represented the feasibility of decoding the imagined speech-based EEG signals. We designed the experimental paradigm strictly to obtain EEG signals with high quality and conducted the experiment in a restricted environment. Moreover, we modified our previous method that the enables increasing the ability of expression by focusing on important features and suppressing unimportant features. Our proposed method indicated the highest average performance for decoding the imagined speech-based EEG signals. We showed that our proposed method increases the ability of expression by focusing on important features and suppresses unimportant features. In other words, our proposed method showed the highest performance, proving that our method has significant strength in learning local features.

In future works, we will develop our proposed model that can analyze the various endogenous BCI paradigms-based EEG signals for the practical BCI systems and can perform with robust classification performances. To this end, we will acquire EEG signals based using new experimental paradigms and will apply various data augmentation methods to solve a lack of data problem.

\section*{ACKNOWLEDGMENT}
The authors would like to thank H.-J. Ahn for designing the experimental paradigm and acquiring the EEG data.

\bibliographystyle{IEEEtran}
\bibliography{REFERENCE}

\end{document}